\documentclass[a4paper]{article}
\usepackage{amsmath, amssymb, graphicx, amsthm, chngpage, verbatim, hyperref, url}
\usepackage[round]{natbib}

\newcommand{\lt}{\left}

\newcommand{\rt}{\right}

\newcommand{\la}{\langle}
\newcommand{\ra}{\rangle}

\newcommand{\uaz}{|\hspace{-2 pt}+\hspace{-2 pt}1_z\ra}
\newcommand{\uazbra}{\la+1_z|}
\newcommand{\daz}{|\hspace{-2 pt}-\hspace{-2 pt}1_z\ra}
\newcommand{\dazbra}{\la-1_z|}
\newcommand{\udaz}{|\hspace{-2 pt}\pm\hspace{-2 pt}1_z\ra}

\newcommand{\uazz}{|\hspace{-2 pt}+\hspace{-2 pt}1Z\ra}
\newcommand{\dazz}{|\hspace{-2 pt}-\hspace{-2 pt}1Z\ra}
\newcommand{\udazz}{|\hspace{-2 pt}\pm\hspace{-2 pt}1Z\ra}

\newcommand{\uax}{|\hspace{-2 pt}+\hspace{-2 pt}1_x\ra}
\newcommand{\dax}{|\hspace{-2 pt}-\hspace{-2 pt}1_x\ra}
\newcommand{\udax}{|\hspace{-2 pt}\pm\hspace{-2 pt}1_x\ra}

\newcommand{\uaxx}{|\hspace{-2 pt}+\hspace{-2 pt}1X\ra}
\newcommand{\daxx}{|\hspace{-2 pt}-\hspace{-2 pt}1X\ra}
\newcommand{\udaxx}{|\hspace{-2 pt}\pm\hspace{-2 pt}1X\ra}

\newcommand{\uay}{|\hspace{-2 pt}+\hspace{-2 pt}1_y\ra}
\newcommand{\dayy}{|\hspace{-2 pt}-\hspace{-2 pt}1_y\ra}

\newcommand{\uai}{|\hspace{-2 pt}+\hspace{-2 pt}1_i\ra}
\newcommand{\dai}{|\hspace{-2 pt}-\hspace{-2 pt}1_i\ra}

\begin{document}
\begin{center}
	{\LARGE \textbf{When Greenberger, Horne and Zeilinger meet Wigner's Friend\\}}
	\vspace{0.5cm}
	Gijs Leegwater\\
	Faculty of Philosophy\\
	Erasmus University Rotterdam\\[0.5cm]
	\today\\[0.5cm]
\end{center}
\begin{abstract}
\noindent A general argument is presented against relativistic, unitary, single-outcome quantum mechanics. This is achieved by combining the Wigner's Friend thought experiment with measurements on a Greenberger--Horne--Zeilinger (GHZ) state, and describing the evolution of the quantum state in various inertial frames. Assuming unitary quantum mechanics and single outcomes, the result is that the Born rule must be violated in some inertial frame: in that frame, outcomes are obtained for which no corresponding term exists in the pre-measurement wavefunction.
\end{abstract}
\section{Introduction}
A central ingredient of quantum mechanics is the unitary evolution of closed systems. The theory also tells us that, after a measurement has been performed upon a system, we can assign a specific pure state to that system: an eigenstate of the measured observable corresponding to the outcome that is obtained, for example $|\!\!\uparrow\ra$. Now suppose such a measurement, say on a particle, takes place inside a laboratory that is a closed system. Describing the measurement process unitarily, the post-measurement state of the laboratory is generally one where the particle ends up being entangled with other parts of the laboratory, like the measurement apparatus, the experimenter, a notebook wherein the results are written down, etc. Quantum mechanics tells us that if there is such entanglement, the particle is not in a pure state; at most we can use a density operator to represent its state. How can this be reconciled with the pure eigenstate $|\!\!\uparrow\ra$ assigned to the particle by the experimenter? Must the unitary evolution of closed systems break down when a measurement takes place inside the system? 

This question lies at the heart of the measurement problem of quantum mechanics. Interpretations of quantum mechanics\footnote{As is common in the literature, with `interpretation of quantum mechanics' we refer not only to interpretations in the literal meaning of the word, but also to alternative theories, like dynamical collapse theories} aim to answer this question by giving a specific account of what happens during a measurement. In some interpretations, the unitary evolution of closed systems is upheld. Such interpretations are examples of what we will call `unitary interpretations' or `unitary quantum mechanics'. Somehow, such interpretations must account for the fact that in the situation given above, the experimenter can assign the pure eigenstate  $|\!\!\uparrow\ra$ to the particle. Indeed, unitary interpretations like Bohmian Mechanics and Many-Worlds provide such an account. However, these two specific interpretations do so at a cost. Bohmian Mechanics introduces a preferred reference frame, which creates tension with relativity theory. The Many-Worlds interpretation introduces an infinity of parallel worlds, prompting various philosophical problems, like how to deal with the probability of measurement outcomes when all possible outcomes actually occur. Of course, for some people, namely Bohmians and Everettians, these costs are not too high. Yet, others would like to see whether a version of unitary quantum mechanics can be found that does not pay such prices. This paper investigates the general possibility of such an interpretation. The answer, of course with some caveats, turns out to be `no': unitary quantum mechanics can only be upheld if either it is denied that measurements have single outcomes, or that the Born rule is violated in some inertial frame.\footnote{Throughout we assume that our background spacetime is Minkowskian.} This result is obtained by considering a thought experiment that combines the `Wigner's Friend' thought experiment with the Greenberger--Horne--Zeilinger (GHZ) state. Put differently, if one wants to maintain that measurements have single outcomes and the Born rule is valid in every inertial frame, it follows that in some cases the unitarity of quantum mechanics breaks down. Various proposals of `dynamical collapse models' have been put forward that specify how unitarity may be violated. Such a violation of unitarity is in principle detectable. Also, the result of the present paper puts constraints on any (future) unitary interpretation of quantum mechanics. Examples of recent attempts are modal interpretations \citep{Dieks1998}, the `Flea on Schr\"odinger's Cat' of \citet{Landsman2013} , and Kent's attempted solution to the `reality problem' (a term he uses for his generalisation of the measurement problem) \citep{Kent2015}.

In the following two sections, we rehearse the Wigner's Friend thought experiment and the Greenberger--Horne--Zeilinger no-go result. Then, we combine the two to reach our main result, presented in Section \ref{SecWFGHZ}. After that, we discuss.

\section{Wigner's Friend}\label{SecWF}
In 1961, Eugene Wigner proposed a thought experiment that focuses on the tension between unitary evolution and wavefunction collapse mentioned in the introduction above \citep{Wigner1961}. The experiment, called `Wigner's Friend', is a variation on Schr\"odinger's Cat. Instead of a cat inside a box, the experiment features a friend of Wigner, whom we call Alice, inside a sealed laboratory. Alice performs a measurement, while Eugene is standing outside.\footnote{To distinguish between Wigner as the author of the 1961 paper and Wigner as the person standing outside the laboratory in the thought experiment, we refer to the latter using his first name Eugene.} After Alice has performed her experiment, Eugene opens the door of the laboratory and asks Alice what the result of her measurement was. The question is now: If Eugene wants to assign a quantum state to the laboratory (including Alice) before he opens the door, what state should this then be? If the state of the laboratory is assumed to have evolved unitarily before opening the door, then, Wigner argued, it will be a macroscopic superposition of states corresponding to different outcomes. This he found unacceptable, because he thought this to imply that his friend is temporarily in a state of `suspended animation', whereas his friend will testify never have been in such a bizarre state. Therefore, Wigner concluded that the correct state for Eugene to assign to the laboratory is a collapsed state, violating unitary evolution. Wigner thought that the presence of human consciousness is the decisive factor in wavefunction collapse, his view is therefore sometimes referred to as `consciousness causes collapse'.

In Wigner's original thought experiment, the measurement of the friend consists of seeing a flash or not. We consider a variation, where Alice measures the $z$-spin of an electron prepared in the $x$-up state. The measurement takes place between times $t_0$ and $t_1$, and we assume the laboratory to be a closed system during the measurement. Describing the measurement as an ideal Von Neumann measurement, we have the pre-measurement state
\begin{align}
|\Psi(t_0)\ra = |\mbox{ready}\ra_L \otimes \uax_A = |\mbox{ready}\ra_L \otimes \sqrt{\frac12} \lt( \uaz_A + \daz_A \rt).\label{WFPsit0}
\end{align}
Here, $A$ is the electron (of which we only consider its spin degree of freedom), while $L$ refers to the rest of the laboratory, including Alice herself. The state $|\mbox{ready}\ra_L$ is a state of the laboratory wherein Alice is just about to perform her measurement, while $\uai$ ($\dai$) is a spin-up (spin-down) state of the electron in the $i$ direction. Note that $\udaz = 1/\sqrt2 (\uax \pm \dax)$. Assuming unitary quantum mechanics for now, the measurement interaction brings about entanglement between the electron and the laboratory:
\begin{align}
|\Psi_\mathrm{uni}(t_1)\ra = \sqrt{\frac12} \lt( |\mbox{Alice $+1_z$}\ra_L \otimes \uaz_A + |\mbox{Alice $-1_z$}\ra_L \otimes \daz_A \rt),\label{WFPureState}
\end{align}
or, expressing the state as a density operator:
\begin{align}
\rho_\mathrm{uni}(t_1) = |\Psi_\mathrm{uni}(t_1)\ra \la \Psi_\mathrm{uni}(t_1) |.
\end{align}
The state $|\mbox{`Alice $+1_z$ ($-1_z$)'}\ra_L$ is the state of a laboratory wherein the measurement apparatus has registered `spin up (down)', and Alice having seen this outcome, etc. Now, if we follow Wigner's line of thought, only one of the terms in \eqref{WFPureState} survives; namely the term that corresponds to the outcome Alice has registered. So, according to Wigner, at $t_1$ we have either
\begin{align}
|\Psi_\mathrm{col}(t_1)\ra &= |\mbox{Alice $+1_z$}\ra_L \otimes \uaz_A, \mbox{ or}\notag\\
|\Psi_\mathrm{col}(t_1)\ra &= |\mbox{Alice $-1_z$}\ra_L \otimes \daz_A.
\end{align}
Since Eugene doesn't know which outcome Alice has obtained, he can use a density operator to represent a mixed state of the laboratory and the electron:\footnote{In the terminology of \citet{Espagnat1971}, this would be an example of a `proper mixture', because the state being a mixture reflects the uncertainty of Eugene about which (pure) state the laboratory and the electron are in.}
\begin{align}
\rho_\mathrm{col}(t_1) = \frac12 \bigl( |\mbox{Alice $+1_z$}\ra_L \la\mbox{Alice $+1_z$}|_L  \otimes \uaz_A \uazbra_A \bigr. \notag\\
\bigl. + |\mbox{Alice $-1_z$}\ra_L \la\mbox{Alice $-1_z$}|_L\otimes \daz_A \dazbra_A \bigr) \label{WFMixedState}
\end{align}
However, we might go against Wigner and assume that for an outside observer as Eugene, the correct state to assign is the unitarily evolved state $\rho_\mathrm{uni}(t_1)$. This is what we call
\begin{quote}
\textbf{Unitary quantum mechanics:} 
As long as a quantum system is closed, i.e. it does not interact with other quantum systems, it evolves unitarily.
\end{quote}
Is there a way for Eugene to establish empirically which of the states 
$\rho_\mathrm{uni}(t_1)$ and $\rho_\mathrm{col}(t_1)$ is the correct one?

If Eugene opens the door of the laboratory and asks his friend Alice what outcome she obtained, this can be considered as a measurement of the laboratory in the basis\footnote{A `measurement in the basis $\{|i\ra\}_i$' is a measurement of an observable that has $\{|i\ra\}_i$ as eigenstates, each with a different eigenvalue.}
\begin{align}
\{ |\mbox{Alice $+1_z$}\ra_L, |\mbox{Alice $-1_z$}\ra_L \}.\label{EigenDoor}
\end{align}
For both of the states $\rho_\mathrm{uni}(t_1)$ and $\rho_\mathrm{col}(t_1)$, the probability of each outcome of this measurement is $50\%$. Therefore, this measurement cannot distinguish between these states.

There are however, in principle, measurements that can distinguish between the states. Define states
\begin{align}
\uazz_{AL} &:= |\mbox{Alice $+1_z$}\ra_L \otimes \uaz_A;\notag\\
\dazz_{AL} &:= |\mbox{Alice $-\hspace{-2 pt}1_z$}\ra_L \otimes \daz_A;\label{Defudazz}\\
\udaxx_{AL} &:= \sqrt{\frac12} ( \uazz \pm \dazz ).\notag
\end{align}
Then, we can rewrite the unitarily evolved state as follows:
\begin{align}
|\Psi_\mathrm{uni}(t_1)\ra &= \sqrt{\frac12} \lt( \uazz_{AL} + \dazz_{AL} \rt)\label{WFPureStateRe}\\
&= \uaxx_{AL}, \mathrm{or}\\
\rho_\mathrm{uni}(t_1) &= \uaxx \la+1X|_{AL}
\end{align}
while the collapsed state can be rewritten as
\begin{align}
\rho_\mathrm{col}(t_1) = \frac12 \Bigl( \uazz \la+1Z|_{AL} + \dazz \la-1Z|_{AL} \Bigr).\label{WFMixedStateRe}
\end{align}
Now consider a measurement of the observable
\begin{align}
\hat{J} := +1 \cdot \uaxx \la+1X|_{AL} -1 \cdot \daxx \la-1X|_{AL}.\label{Aobs}
\end{align}
For this measurement, both outcomes have probability $50\%$ if the state equals $\rho_\mathrm{col}(t_1)$ , while the outcome will with $100\%$ probability be $+1$ if the state equals $\rho_\mathrm{uni}(t_1)$. Therefore, using such a measurement Eugene could in principle determine which is the correct quantum state to assign to the laboratory. \citet{Barr1999}, in discussing Wigner's thought experiment, calls such a measurement a `J-measurement', because he labels his observable (similar to our $\hat{J}$) as $\hat{A}$. Analogously, we will refer to it as a `J-measurement'.

While such a measurement is very hard to perform in reality, as Wigner already noted, quantum mechanics does not forbid such measurements, unless we exploit the fact that some Hermitian operators might not correspond to measurable physical quantities. However, forbidding the J-measurement in this way would seem to be an ad hoc way of avoiding the contradiction presented in this paper.

\section{The Greenberger--Horne--Zeilinger no-go result}\label{SecGHZ}
The Greenberger--Horne--Zeilinger (GHZ) state  was introduced in \citep{GHZ1990} in order to prove the incompatibility of quantum mechanics and local hidden variable theories `without inequalities'. While, in Bell's original theorem, probabilities of (combinations of) measurement outcomes are compared using inequalities, using the GHZ state one only needs to consider which (combinations of) outcomes are possible and which are not.

We give in this section a brief rehearsal of the original GHZ no-go theorem against local hidden variables. This will be helpful, because the contradiction derived in the present paper is similar. Consider three electrons $A$, $B$ and $C$, prepared in the GHZ state\footnote{This GHZ state differs slightly from the original one, having two terms in the $x$-basis instead of the $z$-basis and having an extra factor $\mathrm{i}$ in the second term. This state works just as well for deriving our contradictions and suits us better when combining it with Wigner's Friend.}
\begin{align}
|\mbox{GHZ}\ra := \sqrt{\frac12} ( \uay_A \uay_B \uay_C - \mathrm{i} \dayy_A \dayy_B \dayy_C ).\label{GHZState}
\end{align}
Suppose the spin of each electron is measured in the $x$-direction. To see the possible measurement outcomes, we express the GHZ state in the $x$-basis:\footnote{From this point on, values of summation variables are restricted to $\pm1$.}
\begin{align}
|\mbox{GHZ}\ra = \sum_{a \cdot b \cdot c = -1} e_{abc} |a_x\ra_{A} |b_x\ra_{B} |c_x\ra_{C},\label{GHZinx}
\end{align}
with for all coefficients $|e_{abc}|^2 = 1/4$. Neither the phase nor the exact value of the coefficients concern us, only that they are non-zero so that the terms correspond to possible triples of measurement outcomes. Thus, it turns out that there are four possible triples of outcomes, each with product $-1$.

Similarly, suppose one electron is measured in the $x$-direction, and two electrons are measured in the $z$-direction. The state, expressed in the appropriate bases, equals
\begin{align}
|\mbox{GHZ}\ra &= \sum_{a \cdot v \cdot w = +1} f_{avw} |a_x\ra_A |v_z\ra_B |w_z\ra_C = \sum_{u \cdot b \cdot w = +1} g_{ubw} |u_z\ra_A |b_x\ra_B |w_z\ra_C\notag\\
&= \sum_{u \cdot v \cdot c = +1} h_{uvc} |u_z\ra_A |v_z\ra_B |c_x\ra_C\label{GHZxzz}
\end{align}
with, for all coefficients, $|f_{avw}|^2 = |g_{ubw}|^2 = |h_{uvc}|^2 = 1/4$. It turns out that in these cases, the product of the outcomes always equals $+1$.

We now consider the case where the three electrons are spatially separated and measured simultaneously in the $x$- or $z$-direction. Also, we assume what is usually called local determinism.\footnote{This is just to reproduce the original GHZ result. Local determinism is not assumed to arrive at the main result in Section \ref{SecWFGHZ}.} This means that for individual electron has predetermined values for the measurement outcomes for both the $x$- and the $z$-direction (determinism), and these values are independent of whether the $x$- or the $z$-spin of the other two electrons is measured (locality). Because quantum mechanics itself does not provide for such predetermined values, we introduce a set of hidden variables. Let $a,b,c$ equal the predetermined values for the $x$-direction and $u,v,w$ equal the values for the $z$-direction for electrons $A$, $B$, $C$, respectively; a notation which coincides with the summation variables above. Now, demanding that these predetermined values obey the correlations, predicted by quantum mechanics above, we get the following constraints:
\begin{subequations}
\begin{align}
a \cdot b \cdot c = -1 \label{GHZc1};\\
a \cdot v \cdot w = +1 \label{GHZc2};\\
u \cdot b \cdot w = +1 \label{GHZc3};\\
u \cdot v \cdot c = +1 \label{GHZc4}.
\end{align}
\end{subequations}
By multiplying \eqref{GHZc2}--\eqref{GHZc4} and noticing that $u^2 = v^2 = w^2 = +1$ we get $a \cdot b \cdot c = +1$, in contradiction with \eqref{GHZc1}.

We therefore arrive at the conclusion that there is no local deterministic hidden variable theory compatible with the quantum mechanical predictions for measurements on GHZ states.

\section{Wigner's Friend meets GHZ}\label{SecWFGHZ}
In this section, we will combine the Wigner's Friend thought experiment with the GHZ state. First, we will explicate the assumptions made to arrive at a contradiction. Apart from unitary quantum mechanics, we also assume single outcomes and relativistic quantum mechanics:
\begin{quote}
\textbf{Single outcomes:} Any measurement has only one single outcome. This outcome is independent of the perspective or frame in which the outcome is described.
\end{quote}
We want to emphasise that we acknowledge the strong tension between this assumption and the assumption of unitary quantum mechanics. If the quantum state of a closed system always evolves unitarily, then for any measurement, by considering a system large enough, the wavefunction generally contains terms corresponding to multiple outcomes, not a single outcome. A typical example is the unitarily evolved state \eqref{WFPureState}, which contains a term corresponding to Alice having found the outcome `up' and another term corresponding to her having found the outcome `down'. If the wave function is interpreted realistically, this state seems hard to reconcile with the statement that the measurement only has one of these outcomes. Indeed, it is this difficulty which led Wigner to reject the state \eqref{WFPureState}, and his statement that such a state means that his friend would be in a state of `suspended animation' suggests that Wigner thought that this state is incompatible with his friend having found a single outcome. Yet, as mentioned in the introduction, there are interpretations, such as Bohmian Mechanics and modal interpretations, which aim to reconcile unitary evolution with single outcomes. Moreover, even without picking an existing interpretation of quantum mechanics, many would consider both assumptions desirable features of quantum mechanics. Because the aim of this paper is deriving a no-go result, we do not have to go into the details of a mechanism that explains the reconciliation of unitary evolution and single outcomes, nor need we explain in detail what the necessary and sufficient conditions are for the occurrence of a `measurement'.
Relativistic quantum mechanics is assumed to mean that any inertial frame can be used to describe the evolution of the wavefunction, and that the Born rule holds in every such frame. Here, one must be careful when systems are spatially separated, because of the relativity of simultaneity. When considering a composite system consisting of two or more pointlike systems that are spatially separated, in different frames the hyperplanes of simultaneity intersect the worldlines of the pointlike systems at different spacetime points. This will become clearer once we discuss the thought experiment later in this section. In effect, here we are using the Tomonoga-Schwinger formalism \citep{Tomonaga1946,Schwinger1948}, in a similar way as it is used by \citet{Myrvold2002b}. Note furthermore that in our case, we do not need the full Born rule. We only need the following necessary condition for it: a certain combination of outcomes is possible if and only if, when expressing the pre-measurement wavefunction in the eigenbasis of the measurement, the corresponding term has a non-zero coefficient.

Now, to combine the GHZ state with Wigner's Friend, consider a total of three sealed laboratories, located at the vertices of an equilateral triangle. In addition to laboratory $L$ introduced in the previous section, with Eugene outside and his friend Alice inside, we have one ($M$) with Johnny outside and his friend Bob inside, and one ($N$) with Daniel outside and his friend Charlie inside. Alice, Bob and Charlie again perform a measurement, each on a single electron. But instead of the three electrons being prepared in the $x$-up state, they are prepared in the GHZ state defined in \eqref{GHZState}. Alice, Bob and Charlie perform their measurements in the $z$-direction between times $t_0$ and $t_1$. Expressed in the $z$-basis, the GHZ state reads
%
\begin{align}
|\mbox{GHZ}\ra = \sum_{u,v,w=\pm1} d_{uvw} |u_z\ra_A |v_z\ra_B |w_z\ra_C,\label{GHZzzz}
\end{align}
with for all coefficients $|d_{uvw}|^2 = 1/8$. So, all triples of outcomes are possible. Now, assume that the measurements are ideal von Neumann measurements and define $\udazz_{BM}$ and $\udazz_{CN}$ for Bob's and Charlie's laboratories, and their electrons, in the same way as we have defined $\udazz_{AL}$ in \eqref{Defudazz}. For example, we have
\begin{align}
\uazz_{BM} &:= |\mbox{Bob $+1_z$}\ra_M \otimes \uaz_B;\\
\dazz_{BM} &:= |\mbox{Bob $-1_z$}\ra_M \otimes \daz_B.
\end{align}

The pre-measurement state is now
\begin{align}
|\Phi(t_0)\ra = |\mbox{ready}\ra_L |\mbox{ready}\ra_M |\mbox{ready}\ra_N \otimes |\mbox{GHZ}\ra_{ABC},
\end{align}
where, as in \eqref{WFPsit0}, $|\mbox{ready}\ra_{L/M/N}$ are the pre-measurement states of the laboratories of Alice, Bob and Charlie respectively. Just as Alice's measurement, in the introduction to Wigner's Friend above, took the state from \eqref{WFPsit0} ($|\Psi(t_0)\ra$) to \eqref{WFPureState} ($|\Psi_\mathrm{uni}(t_1)\ra$), the state of the laboratories and electrons after the measurements is now
\begin{align}
|\Phi(t_1)\ra = \sum_{u,v,w=\pm1} d_{uvw} |uZ\ra_{AL} |vZ\ra_{BM} |wZ\ra_{CN}.
\end{align}
This state has the same form as the initial GHZ state \eqref{GHZzzz}, but with the states $|u_z\ra_A$, $|v_z\ra_B$, $|w_z\ra_C$ being replaced with $|uZ\ra_{AL}$, $|vZ\ra_{BM}$, $|wZ\ra_{CN}$. Now, analogously to \eqref{Defudazz}, defining the states
\begin{align}
\udaxx := \sqrt{\frac12} ( \uazz \pm \dazz ),
\end{align}
for the systems $BM$ and $CN$, and noting that $\udaxx$ and $\udazz$ are related in the same way as $\udax$ and $\udaz$, then, using \eqref{GHZinx}, we can write the total state at $t_1$ as
\begin{align}
|\Phi(t_1)\ra = \sum_{a \cdot b \cdot c=-1} e_{abc} |aX\ra_{AL} |bX\ra_{BM} |cX\ra_{CN},\label{Psit1}
\end{align}
with again the coefficients satisfying $|e_{abc}|^2 = 1/4$. Then, between $t_1$ and $t_2$, Eugene performs a J-measurement, that is, he performs a measurement of the observable $\hat{J}$, defined in \eqref{Aobs}. Simultaneously, Johnny and Daniel perform measurements of the similarly defined observables
\begin{align}
\hat{K} &:= +1 \cdot \uaxx \la+1X|_{BM} -1 \cdot \daxx \la-1X|_{BM};\notag\\
\hat{M} &:= +1 \cdot \uaxx \la+1X|_{CN} -1 \cdot \daxx \la-1X|_{CN}.
\end{align}
Just as with measurements in the $x$-direction on the original GHZ state, we see that the product of the outcomes equals $-1$.
\begin{figure}
\includegraphics[width=\textwidth]{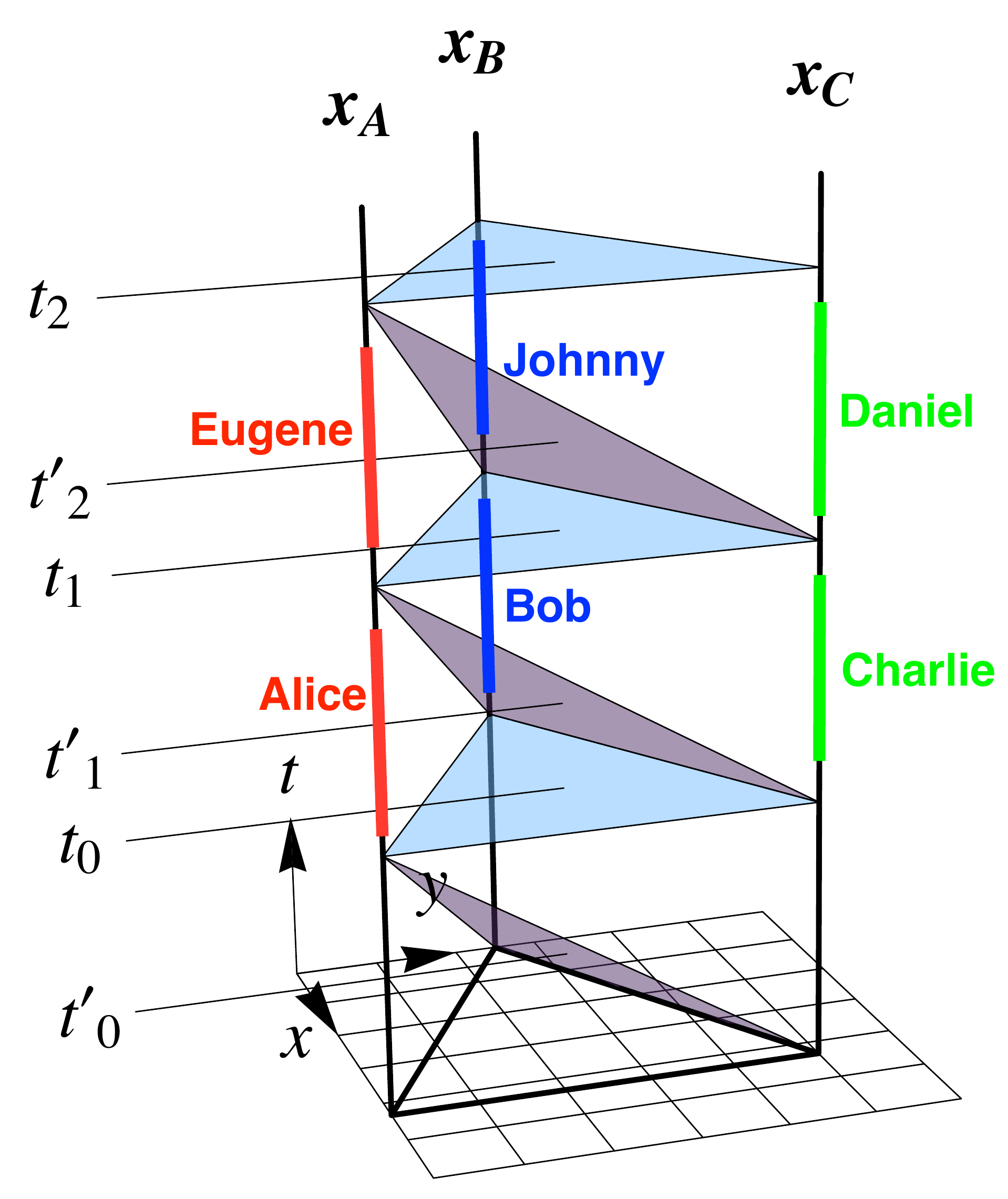}
\caption{In the reference frame $\Sigma$ (unprimed coordinates), $ABC$ is prepared in the GHZ state at $t_0$. The measurements of Alice, Bob and Charlie take place simultaneously and at locations $\mathbf{x}_A$, $\mathbf{x}_B$ and $\mathbf{x}_C$ (which form an equilateral triangle), between $t_0$ and $t_1$. The measurements of Eugene, Johnny and Daniel take place, also at $\mathbf{x}_A$, $\mathbf{x}_B$ and $\mathbf{x}_C$ and simultaneously in $\Sigma$, between $t_1$ and $t_2$, where $t_2 - t_1 = t_1 - t_0$. Also, the spacetime point $(t_1, \mathbf{x}_A)$ is spacelike separated from $(t_0, \mathbf{x}_B)$, i.e., setting $c=1$, $t_1 - t_0 < ||\mathbf{x}_A - \mathbf{x}_B||$ (this ensures, together with the above conditions, spacelike separation between any two measurements at different laboratories). The frame where $(t_1, \mathbf{x}_A)$ is simultaneous with $(t_0, \mathbf{x}_B)$ and $(t_0, \mathbf{x}_C)$ is defined as $\Sigma'$ (primed coordinates); the hyperplane of simultaneity containing these points has $t'_1$ as its time coordinate. The frames $\Sigma''$ and $\Sigma'''$ (not displayed here) are defined similarly, by cyclicly permuting the triplet $\la A,B,C \ra$: In $\Sigma''$, Johnny's measurement is simultaneous with Alice's and Charlie's, while in $\Sigma'''$, Daniel's measurement is simultaneous with Alice's and Bob's. The laboratories are, for simplicity, assumed to be pointlike. This may look like a bold simplification, but note that the distance between the laboratories can be made arbitrarily large. For the same reason, the relative velocity between the frames can be made arbitrarily small by choosing $||\mathbf{x}_A - \mathbf{x}_B|| \gg t_1 - t_0$.}
\label{Fig1}
\end{figure}

Now suppose the two measurements at each laboratory take place at spacelike separation from the four measurements at the other laboratories (see Figure \ref{Fig1} for details). Then we can choose another frame $\Sigma'$, such that first Alice performs her measurement, then Eugene, Bob and Charlie perform their measurements simultaneously, and then Johnny and Daniel perform their measurements. For simplicity, we assume that, before $t_0$, the states of the electrons and the laboratories have been constant for a while. Then, before Alice's measurement, the state of her laboratory and the electrons is
\begin{align}
|\Phi'(t'_0)\ra := \sum_{a \cdot v \cdot w = +1} f_{avw} |\mbox{ready'}\ra_L |a'_x\ra_{A} |v'_z\ra_B |w'_z\ra_C.\label{Psit'0}
\end{align}
The primed states in frame $\Sigma'$ are, following Wigner's Theorem \citep{Wigner1959}, related to the unprimed states in $\Sigma$ by a unitary operator. Then, between $t'_0$ and $t'_1$, Alice performs her measurement. The only nontrivial evolution that takes place in this period is that of Alice's laboratory and her electron. This results in the state
\begin{align}
|\Phi'(t'_1)\ra = \sum_{a \cdot v \cdot w = +1} f_{avw} |aX'\ra_{AL} |v'_z\ra_B |w'_z\ra_C.\label{Psit'1}
\end{align}
Note that also this state has the same form as the original GHZ state \eqref{GHZxzz}, but with $|a'_x\ra_A$ replaced by $|aX'\ra_{AL}$. Now, between $t'_1$ and $t'_2$, Eugene performs a J-measurement, while Bob and Charlie perform a simple experiment of the $z$-spin of electrons $B$ and $C$. The state \eqref{Psit'1} is here expressed in the corresponding measurement bases, so we can directly infer that the product of the outcomes always equals $+1$.

Likewise, there is a frame $\Sigma''$ wherein Johnny measures simultaneously with Alice and Charlie, and a frame $\Sigma'''$ wherein Daniel measures simultaneously with Alice and Bob. In these frames, the states just before the measurements, expressed in the measurement bases, are
\begin{align}
|\Phi''(t''_1)\ra &= \sum_{u \cdot b \cdot w = +1} g_{ubw} |u''_z\ra_A |bX''\ra_{BM} |w''_z\ra_C;\label{Psit''1}\\
|\Phi'''(t'''_1)\ra &= \sum_{u \cdot v \cdot c = +1} h_{uvc} |u'''_z\ra_A |v'''_z\ra_B |cX'''\ra_{CN};\label{Psit'''1}
\end{align}
Labelling the outcomes of Alice, Bob, Charlie, Wigner, Johnny and Daniel as $a, b, c, u, v, w$ respectively, we get from \eqref{Psit1}--\eqref{Psit'''1} the four GHZ constraints:
\begin{subequations}
\begin{align}
a \cdot b \cdot c = -1 \label{GHZc1a}\\
u \cdot b \cdot c = +1 \label{GHZc2a}\\
a \cdot v \cdot c = +1 \label{GHZc3a}\\
a \cdot b \cdot w = +1 \label{GHZc4a}
\end{align}
\end{subequations}
There are exactly the constraints \eqref{GHZc1a}--\eqref{GHZc4a}, which were shown to be contradictory. We therefore arrive at the conclusion that, assuming single outcomes, not all outcomes can be as predicted by unitary quantum mechanics, in every reference frame. In some frame, there must be a combination of outcomes for which there is no corresponding term in the pre-measurement state.
\section{Discussion}
\subsection{Non-ideal measurements}
In the sections above we have assumed that the measurements are ideal Von Neumann measurements. This ensures that the post-measurement states have a simple form, and that the measurements of Eugene, Johnny and Daniel are exactly like the A-measurements as discussed by Barrett. However, this assumption can be dropped, as long as the measurements can be represented by \emph{some} unitary transformation on the laboratories and the electrons, which is the case if unitary quantum mechanics is assumed. In particular, if these transformations are represented by operators $U_{AL}$, $U_{BM}$ and $U_{CN}$, then $\uazz_{AL}$ and $\dazz_{AL}$, defined in \eqref{Defudazz}, can be redefined as
\begin{align}
\uazz_{AL} &:= U_{AL} (|\mbox{`ready'}\ra_L \otimes \uaz_A);\\
\dazz_{AL} &:= U_{AL} (|\mbox{`ready'}\ra_L \otimes \daz_A),
\end{align}
and similar redefinitions for $\udazz_{BM}$ and $\udazz_{CN}$. The fact that unitary transformations take orthogonal states to orthogonal states ensures that the states defined above are still orthogonal. The derivation can then be repeated using these redefinitions, and the contradiction follows as before. Note that the redefinition works through in other definitions. For example, the observable $\hat{J}$ is defined in terms of $\udaxx$, which in turn are defined in terms of the redefined $\udazz$.

Using these redefinitions, the steps in Section \ref{SecWFGHZ} can be applied to derive a contradiction also for interpretations that try to avoid the measurement problem by rejecting the standard Von Neumann measurement scheme to avoid macroscopic entanglement while still retaining unitary quantum mechanics. An example of this is the `Flea on Schr\"odinger's Cat' interpretation of \citet{Landsman2013}. In that interpretation, a perturbation of the Hamiltonian allegedly makes sure that no macroscopic superposition results from a measurement. Such an interpretation should however deal with the case presented in this article. While rejecting the standard Von Neumann measurement account; a measurement still corresponds to \emph{some} unitary evolution in Landsman's `Flea' approach. Defining $U_{AL}$, $U_{BM}$ and $U_{CN}$ to represent these evolutions, the contradiction can also be derived for Landsman's approach. While it might be very hard in practice for Eugene and Johnny to perform the measurements of $\hat{J}$ and $\hat{K}$ in bases involving these unitary transformations, which include the `flea' perturbations, there is no apparent reason why such measurements would be impossible in principle. And in that case, we seem again to have a case where the Born rule is violated in some inertial frame.
\subsection{Interpretations of quantum mechanics that avoid the contradiction}
The main result of this paper is a negative one: it tells us what kind of interpretation of quantum mechanics is \emph{not} possible. However, this might lead the way to a positive result: what kind of interpretation of quantum mechanics \emph{is} possible?

\subsubsection{Preferred reference frame}
One route to go is to assume a preferred reference frame in which the Born rule is valid while it may be violated in other reference frames. For example, if $\Sigma$ is the preferred reference frame, then the outcomes of both the joint measurements of Alice, Bob and Charlie and that of Eugene, Johnny and Daniel obey the GHZ correlations, but the constraints \eqref{GHZc2a}--\eqref{GHZc4a} cannot be derived anymore, since in the frame $\Sigma$ the quantum state never has the form \eqref{Psit'1}--\eqref{Psit'''1}.

More generally, if three measurements take place simultaneously in the preferred reference frame, then the corresponding constraint from \eqref{GHZc1a}--\eqref{GHZc4a} is satisfied. This implies that at least one of the other three constraints is violated, and such a violation therefore implies that the measurements corresponding to the outcomes appearing in the violated constraint did not take place simultaneously in the preferred frame.
Does this mean we can empirically establish what is (not) the preferred reference frame, by comparing the outcomes of the measurements and seeing which of the GHZ constraints is violated?

Unfortunately, this does not seem to be possible. Suppose Eugene wants to collect all six measurement outcomes. Of course, he will have no problem knowing the outcome of his own measurement. Neither will he have issues with asking Johnny and Daniel for their outcomes. But he will have a harder time retrieving the outcomes of Alice, Bob and Charlie. Take Alice's outcome. Depending on Eugene's outcome, he may assign one of the eigenstates of the J-measurement ($\uaxx$) to Alice's and laboratory and her particle. By the relations given in \eqref{Defudazz}, each of these eigenstates is a superposition of two states: one state representing Alice having found `up', and one state representing Alice having found `down'. So, if Alice is asked what outcome she obtained, there is equal probability of her saying up as saying down, and there is no guarantee that what she says corresponds to the outcome she actually obtained during her own measurement. Eugene's measurement effectively `erases' Alice's outcome. This will become even clearer in the example discussed in Section \ref{FeasA}.


Bohmian Mechanics is the foremost example of an interpretation that includes a preferred reference frame. In that frame, the distribution of particle positions is given by the squared modulus of the wavefunction over configuration space. In any other reference frame than the preferred frame, the particle positions may not be distributed according to the wavefunction in that (non-preferred) frame. In fact, in general it is not possible for the particle positions to be distributed according to the wavefunction in every frame \citep{Berndl1995}. In this sense, the preferred reference frame manifests itself at the level of particle positions. The preferred reference frame can, however, not be detected because the particle positions are not directly accessible. What is surprising about the result of the present paper is that the preferred reference frame manifests itself also at the level of measurement outcomes. However, as mentioned before, also in this case the frame cannot be detected, since not all six outcomes can be brought together and compared.

\subsubsection{Many Worlds}
Another possible way out would be to deny that there are single outcomes. Assuming single outcomes allowed us to assign single values to the variables $a,b,c,u,v,w$ and consider the fixed correlations between these variables. For all of the measurements discussed above, there are two possible outcomes: $+1$ and $-1$, each with a $50\%$ probability. This means that if we look at the branching structure of the unitarily evolving wavefunction, for every individual measurement outcome, there is a branch containing that outcome. This seems to forbid us to assigning single values to the variables $a,b,c,u,v,w$.

It would be interesting to investigate how exactly the thought experiment in this paper would play out in specific versions of many-worlds quantum mechanics. For example, in the `divergence' view of many-worlds, worlds do not split, and there is a fixed number of them. If that is the case, it seems that in every one of those worlds, there are six single outcomes for the measurements, and therefore in any one of those worlds the contradiction can again be derived. However, working out the details of this, as well as considering other versions of many-worlds, falls outside the scope of this paper.

\subsubsection{Kent's proposal}
There might be other ways to evade the contradiction. \citep{Kent2015} has recently proposed an interpretation that seems to fall in the category of interpretations of to quantum mechanics targeted in this article: relativistic, single-outcome and unitary. In Kent's interpretation, additional to the unitarily evolving wavefunction there is a boundary condition consisting of determinate values of mass-energy along some future hypersurface. Using this boundary condition one can calculate the stress-energy at every point in spacetime, and this ensures that there are single outcomes. However, this only works for measurement outcomes for which there exist a record on this future hypersurface, i.e. different measurement outcomes must correspond to different mass-energy configurations on the final hypersurface. In the argument presented in this article, the measurements by Alice, Bob and Charlie are contained within an isolated laboratory, and, as explicated in the following subsection, the records of their outcomes are erased by the measurements of Eugene, Johnny and Daniel. Therefore, in Kent's interpretation, the first measurements have no single outcomes, and in this way the contradiction is evaded. Whether Kent's interpretation can lead to a full-fledged, satisfactory account of quantum mechanics needs more investigation. 



\subsection{The feasibility of J-measurements}\label{FeasA}
We have deliberately presented a thought experiment in which ordinary quantum experiments, combined with the familiar J-measurements, are enough to arrive at a contradiction. However, while the eigenstates of the J-measurements are easy to write down, these measurements are hard, if not impossible, to perform in practice. We have already mentioned that such measurements in effect erase the previous measurements result; they must get rid of all traces from which one can infer the outcome of the first measurement. To see how peculiar these measurements are, consider what happens when they are performed on a laboratory starting in a definite state $|\pm1Z\ra_{AL}$ of containing Alice, who has found $z$-spin up (starting with a electron prepared in the $z$-up state instead of $x$-up). Considering the eigenstates of the J-measurement $\udaxx = 1/\sqrt2 ( \uazz \pm \dazz )$, we see each outcome has probability $1/2$. However, both outcomes leave the system in an eigenstate that also contains a term corresponding to a laboratory where Alice has found $z$-spin \emph{down}. So, if subsequently the simple measurement is performed of `opening the door' (with eigenstates \eqref{EigenDoor}), then there is a probability $1/2$ of Alice telling that she had outcome \emph{down}, and also finding evidence of this in the laboratory (there might be a computer which has the outcome in its memory, the result may be printed on paper etc.). So, the J-measurement can effectively change a laboratory from a definite state of containing an experimenter who found `up' to a state of containing an experimenter who found `down', illustrating the complexity of such measurements.

The attractiveness of unitary quantum mechanics is that a measurement interaction is treated as any other interaction. If we are going to forbid operations that undo these interactions, then we seem to have gone back to granting a special status to measurements, because they will have become fundamentally irreversible processes (see also \citet{Brukner2017}). One might wonder what the point of considering unitary quantum mechanics is, if measurements that distinguish it from collapse quantum mechanics are fundamentally forbidden.

\subsection{Myrvold's no-go result for relativistic modal interpretations}
The result of this paper is similar to that of \citet{Myrvold2002}, where a no-go theorem is presented for modal interpretations exhibiting `serious' Lorentz invariance. This no-go theorem applies to modal interpretations where local definite properties corresponding to some fixed observable $R$ are assigned to systems. These properties are represented by eigenvalues of the fixed observable. The demand of `serious' Lorentz invariance then requires that the `Relativistic Born Rule' is satisfied.\footnote{This rule, although bearing the same name, differs from the relativistic Born rule defined in the present paper.} A necessary condition for the satisfaction of Myrvold's Relativistic Born Rule is that along all spacelike hyperplanes, the projection of the quantum state along that hyperplane onto the subspaces corresponding to the possessed properties is nonzero. Then by using a Hardy state \citep{Hardy1992}, an example of the evolution of two systems is provided where the Relativistic Born Rule cannot be satisfied.

There are some important differences between the present result and that of Myrvold. First, the present result aims to be more general. It is not only aimed at modal interpretations with properties for a fixed observable, but also at other single-world, unitary interpretations of quantum mechanics such as Kent's proposal and Landsman's `Flea', and possible future proposals. While Myrvold's Relativistic Born Rule concerns (possibly unobserved) possessed definite properties, we only consider actually observed measurement outcomes. The current result can therefor be seen as being similar to that of Myrvold, but with the possessed definite properties `elevated' to measurement outcomes. Denying the existence of (single) measurement outcomes seems much harder than denying the existence of possessed definite properties, making the current result more general.

Regarding the content of the result, a difference is that Myrvold uses a Hardy state while in the present paper a GHZ state is used. Actually, the result in the present paper could also have been achieved using a Hardy state, or a Bell state. The advantage of considering a Hardy or GHZ state instead of a Bell state is that no probabilities have to be considered, only which combinations of outcomes are and aren't possible. The advantage of using a Hardy state compared to using a GHZ state is that only two parties have to be considered, instead of three as in the current paper. However, using the GHZ state also has an advantage: the contradiction becomes apparent in every run of the thought experiment, while for the Hardy state, the contradiction only becomes apparent for some runs of the thought experiment. In more detail: when the Hardy state would have been used in the current article, there would be only four instead of six outcomes. Now, quantum mechanics predicts, with nonzero (but not unity) probability, values for some of these outcomes that are incompatible with any possible values for the rest of the outcomes, resulting in a contradiction. Whether one prefers the Hardy state, which has less parties to deal with, or the GHZ state, which results in a contradiction on every single run, is largely a matter of taste.

\subsection{Frauchiger \& Renner's no-go result for consistent single-world quantum mechanics}
Another result to which the current paper is similar is the recent no-go theorem by \citet{Frauchiger2018} (F\&R). F\&R claim to arrive at a contradiction for single-world quantum mechanics even without considering relativity theory; their result is derived using only a single reference frame. F\&R arrive at a contradiction starting from three assumptions called Single-World, Quantum Mechanics and Consistency. Like Myrvold, F\&R use a Hardy state to arrive at their conclusion, but it seems they could equivalently have used a Bell state or a GHZ state. It would take too far to examine the exact differences between the two results. However, we can say that in our opinion a crucial difference lies in the fact that F\&R do not seem to assume unitary quantum mechanics in the way we do. While we assume, that when predicting measurement outcomes, the system is supposed to have evolved unitarily before the measurement (even when a measurement takes place inside the system), F\&R seem to mix unitary quantum mechanics with collapse in a peculiar way. See also \citet{Baumann2016} and \citet{Sudbery2017} for critical views on F\&R's result.

\section{Acknowledgements}
\noindent I would like to thank F.A. Muller, Jeremy Butterfield, Renato Renner and Guido Bacciagaluppi for valuable discussions and corrections. This work is part of the research programme `The Structure of Reality and the Reality of Structure', which is financed by the Netherlands Organisation for Scientific Research (NWO).

\bibliographystyle{model5-names}
\bibliography{refs}

\end{document}